\documentclass[sn-mathphys,Numbered]{sn-jnl}

\usepackage{subcaption}

\usepackage{graphicx}%
\usepackage{multirow}%
\usepackage{amsmath,amssymb,amsfonts}%
\usepackage{amsthm}%
\usepackage{mathrsfs}%
\usepackage[title]{appendix}%
\usepackage{xcolor}%
\usepackage{textcomp}%
\usepackage{manyfoot}%
\usepackage{booktabs}%
\usepackage{algorithm}%
\usepackage{algorithmicx}%
\usepackage{algpseudocode}%
\usepackage{listings}%

\theoremstyle{thmstyleone}%
\theoremstyle{thmstyletwo}%

\theoremstyle{thmstylethree}%

\begin{document}

\title[Apodized Slanted Teeth Grating Couplers for LiDAR Applications]{ Apodized Slanted Teeth Grating Couplers for LiDAR Applications}

\author[1]{\fnm{Vahram} \sur{Voskerchyan}}\email{vahram.voskerchyan3@uvigo.gal}

\author[1]{\fnm{Yu} \sur{Tian}}\email{yu.tian@uvigo.gal}

\author[2]{\fnm{Francisco M.} \sur{Soares}}\email{chicomsoares@live.com}

\author[1]{\fnm{David Álvarez} \sur{Outerelo}}\email{dalvarez@com.uvigo.es}

\author*[1]{\fnm{Francisco J.} \sur{Diaz-Otero}}\email{fjdiaz@com.uvigo.es}

\affil*[1]{\orgdiv{Telecommunication}, \orgname{University of Vigo}, \orgaddress{\street{Rua Maxwell s/n}, \city{Vigo}, \postcode{36310 },  \country{Spain}}}

\affil[2]{\orgname{Soares Photonics}, \orgaddress{\city{Lisbon}, \country{Portugal}}}


\abstract{Solid state LiDAR systems traditionally rely on costly active components for efficient beam scanning. In this study, we propose a cost-effective, purely passive steering approach using apodized slanted grating couplers. Through apodization, we achieve a uniform upward emission profile and enhanced upward transmission. Theoretical calculations indicate successful steering of 91.5$^\circ$x42.8$^\circ$. Experimental results closely match theoretical predictions, validating the capabilities of our passive steering concept. Additionally, the grating couplers, with a length of 3mm, enable a farfield FWHM of 0.026$^\circ$, further enhancing sensing resolution.}




\maketitle

\section{Introduction}\label{sec1}
Solid-state LiDAR systems play a crucial role as a key technology in various applications \cite{Doylend2020}. Their compact design, which avoids bulkiness, makes them particularly appealing for seamless integration into devices such as drones and self-driving cars. This characteristic holds the potential to drive innovation in sectors like transportation, precision agriculture, and urban planning. Due to the rising need for LiDAR systems
optical beam formation and scanning techniques are currently in high demand.

Several method for beam scanning, including VCELS\cite{Moench2016}, MEMS\cite{Wang2DBeamsteering}, edge fire arrays \cite{EndFirePhasedArray2017} and optical phased arrays (OPAs)\cite{Poulton2017}, have been developed for LiDAR systems. OPAs, in particular, have garnered significant attention due to their capability to achieve a low divergence angle, high output power, and low steering power. Silicon photonic technology is compatible with complementary metal-oxide-semiconductor (CMOS) technology facilitates the large-scale production of these devices. This advantage allows for the creation of precise and cost-effective LiDAR systems that can be mass-produced\cite{Michaelwatts}.

What sets OPAs apart is their ability to operate with low energy consumption, positioning them as sustainable and efficient solutions \cite{VanAcoleyen2010}.

Grating couplers serve as a fundamental building block in optical phased arrays \cite{Guo2021}. When arranged linearly, the emitted light from these couplers undergoes constructive interference. By manipulating the phase of the waveguide, steering can be achieved. In silicon photonics thermo-optical phase modulators are commonly employed for steering \cite{Acoleyen2009}. However, this can complicate the fabrication process of devices with densely integrated grating coupler antennas, leading to increased production costs. The integration of active components on top of the fabricated gratings introduces additional complexities. For instance, thermo-optical components, commonly employed for active beam steering, can contribute to device heating, posing challenges, particularly in applications with stringent thermal constraints, such as space-based applications. 

In this paper we introduce a novel apodized slanted grating coupler that can potentially reduce the costs of the fabrication of the LiDAR systems while also improving the performance by having large Field of View and a smallar spot size due to the dimensions of the single transceiver. The incorporation of slanted gratings in a linear arrangement of gradually-tilted grating couplers facilitates simultaneous illumination and reception of light from all angular directions. This passive configuration eliminates the necessity for optical phased arrays or numerous individually-controlled phase shifters, streamlining the design for efficient 3D environmental mapping \cite{Voskerchyan2021}.

\section{Design of the Slanted Grating Coupler}
\begin{figure}[h]%
\centering
\includegraphics[width=0.3\textwidth]{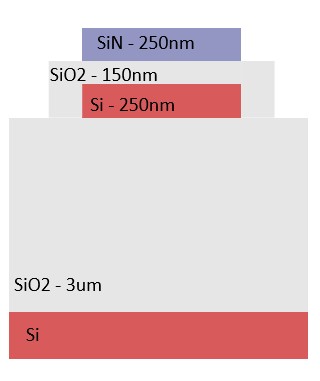}
\caption{The layer stack of the fabricated device and the thicknesses of each material layer.}\label{fig:layerstack}
\end{figure}

Recent works in OPA design have shown that a double-layer structure \cite{Wang2023} was chosen to increase the emitting strength of the grating in the upward direction. The material layerstack is shown in the figure \ref{fig:layerstack}. 

In the FDTD simulation software, we conducted a comprehensive sweep of the layer thickness to enhance the contrast between upward and downward transmissions, aiming to maximize the upward light while maintaining output light. 

\begin{figure}[h]%
\centering
\includegraphics[width=0.5\textwidth]{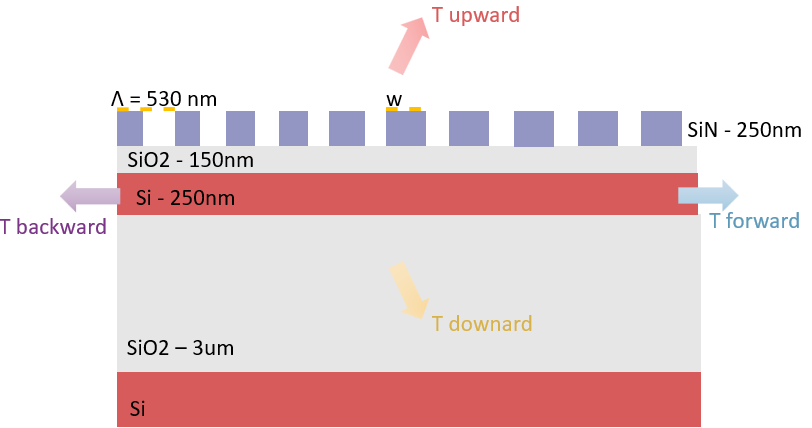}
\caption{The side view of the grating coupler showing the layer thickness and directions of the transmissions.}\label{fig:side_view}
\end{figure}

The schematics of the  apodized grating is shown in the figure \ref{fig:side_view}.The arrows show the directions of the light propagation inside the grating coupler. The period is chosen to be $\Lambda = 0.53\mu$m so when the light radiates at 1550nm it is centered around 0$^\circ$. The $w$ is the duty cycles of the grating which represents the width of the grating teeth.

\begin{figure}[h]%
\centering
\includegraphics[width=0.5\textwidth]{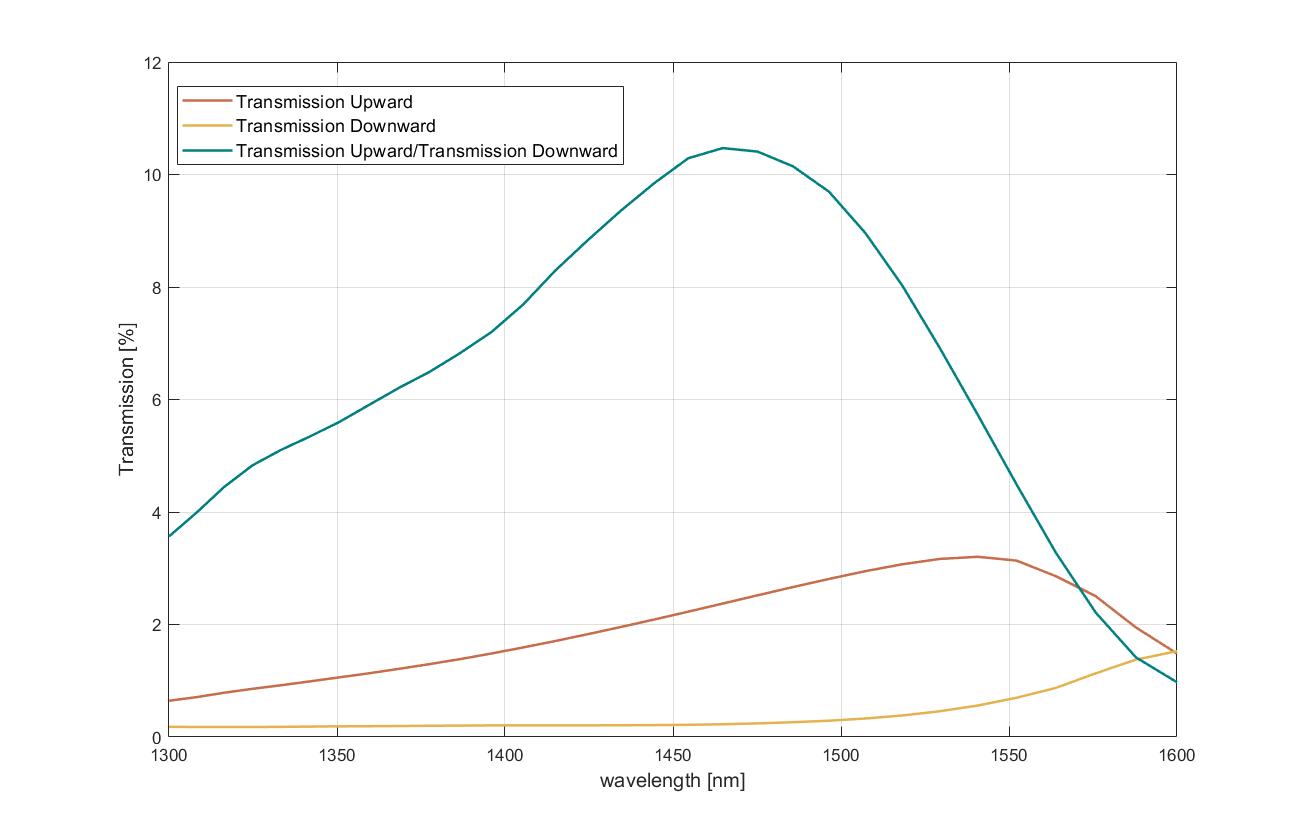}
\caption{The downward and upward transmission difference.}\label{fig:transmission_difference}
\end{figure}

With SiO2 layer thickness set to 150 nm between the silicon (Si) and silicon nitride (SiN) layers, and Si-SiN layer thicknesses maintained at 250 nm each, the calculated difference between the upward and downward transmissions (T upward/T downward) is 10.4 at the $\lambda$ = 1464 nm. The provided graph (\ref{fig:transmission_difference} depicts the transmission contrast across wavelengths spanning from 1.3 $\mu$m to 1.6 $\mu$m, making the grating useful for broadband region. This difference is computed using a filling factor of 0.5, and the grating structure has 42 teeth, the filling factor is the ration between the period $\Lambda$ and duty cycle $w$ which is expressed as $ff=\frac{\Lambda}{w}$.

After establishing the thicknesses of the layer stack, simulations for the radiation angles emanating from the grating device have been conducted. 

\begin{figure}[h]%
\centering
\includegraphics[width=0.5\textwidth]{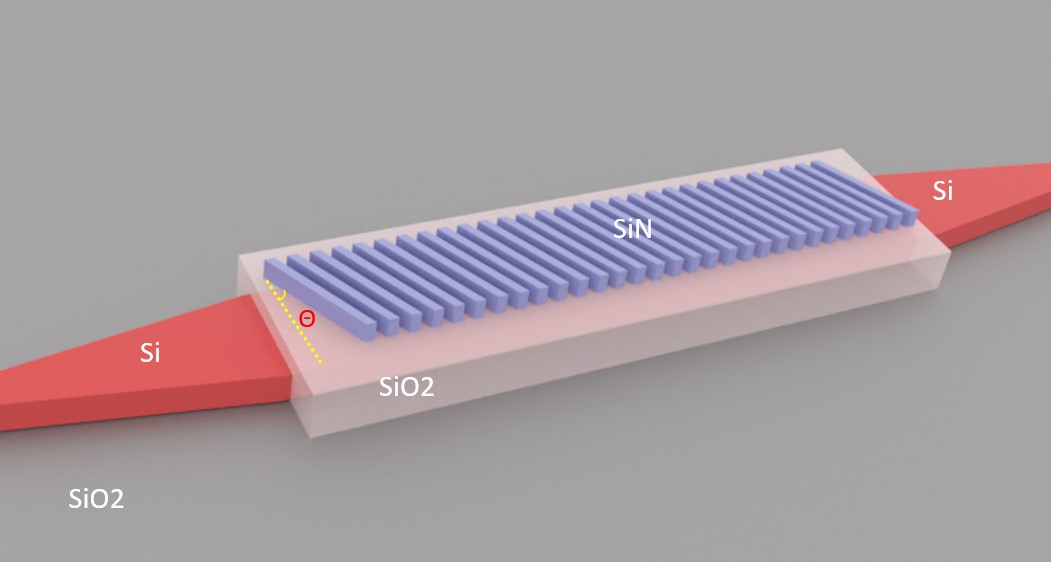}
\caption{Conceptual image of the grating with the $\Theta$ slanting angle.}\label{fig:3dconceptual}
\end{figure}

The provided image (figure \ref{fig:3dconceptual}) illustrates the conceptual design of the grating coupler, where $\Theta$ represents the slanting angle of the grating.

\begin{figure}
\centering
\begin{subfigure}{.5\textwidth}
  \centering
  \includegraphics[width=.7\linewidth]{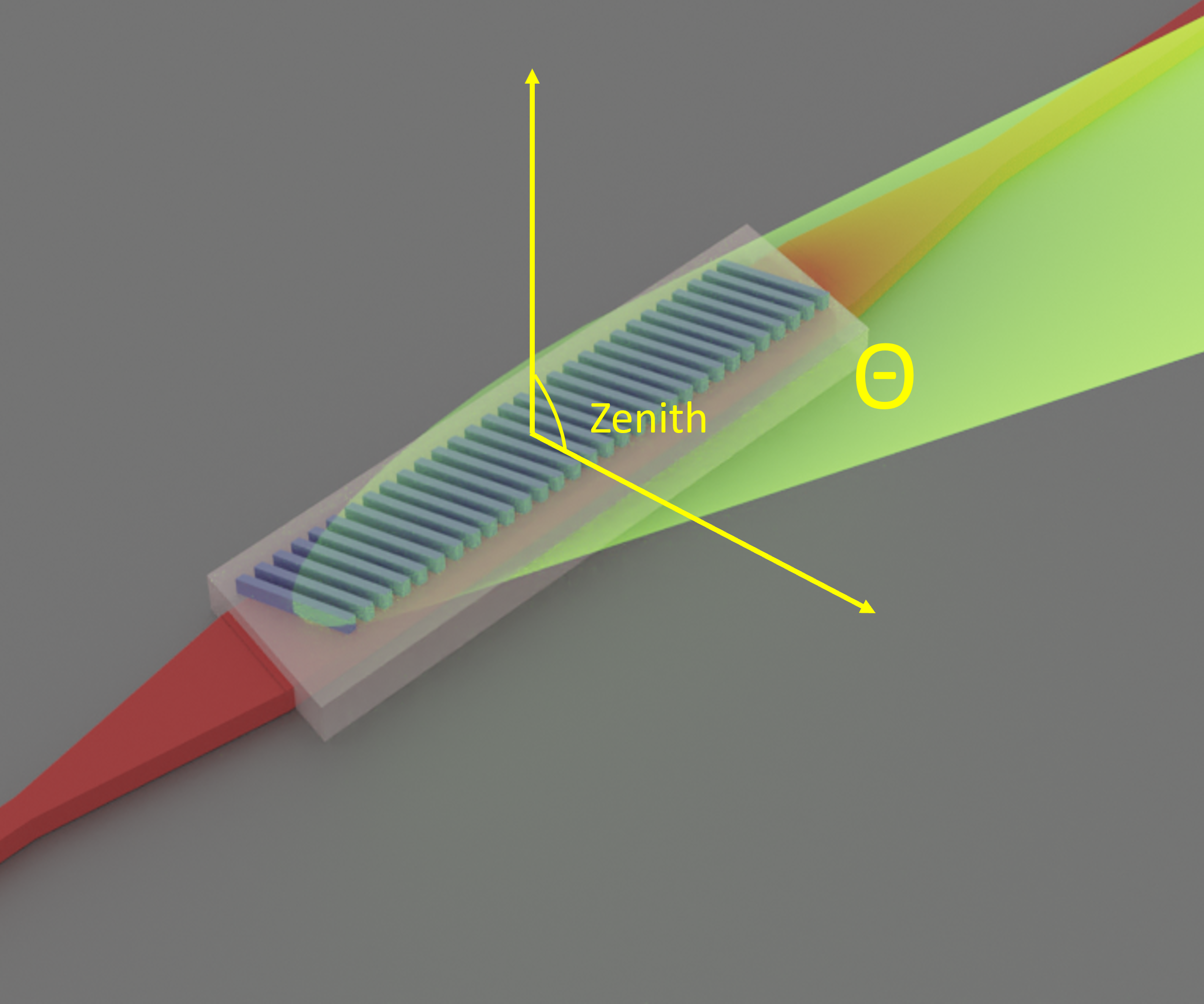}
  \caption{}
  \label{fig:3dslant:a}
\end{subfigure}%
\begin{subfigure}{.5\textwidth}
  \centering
  \includegraphics[width=.7\linewidth]{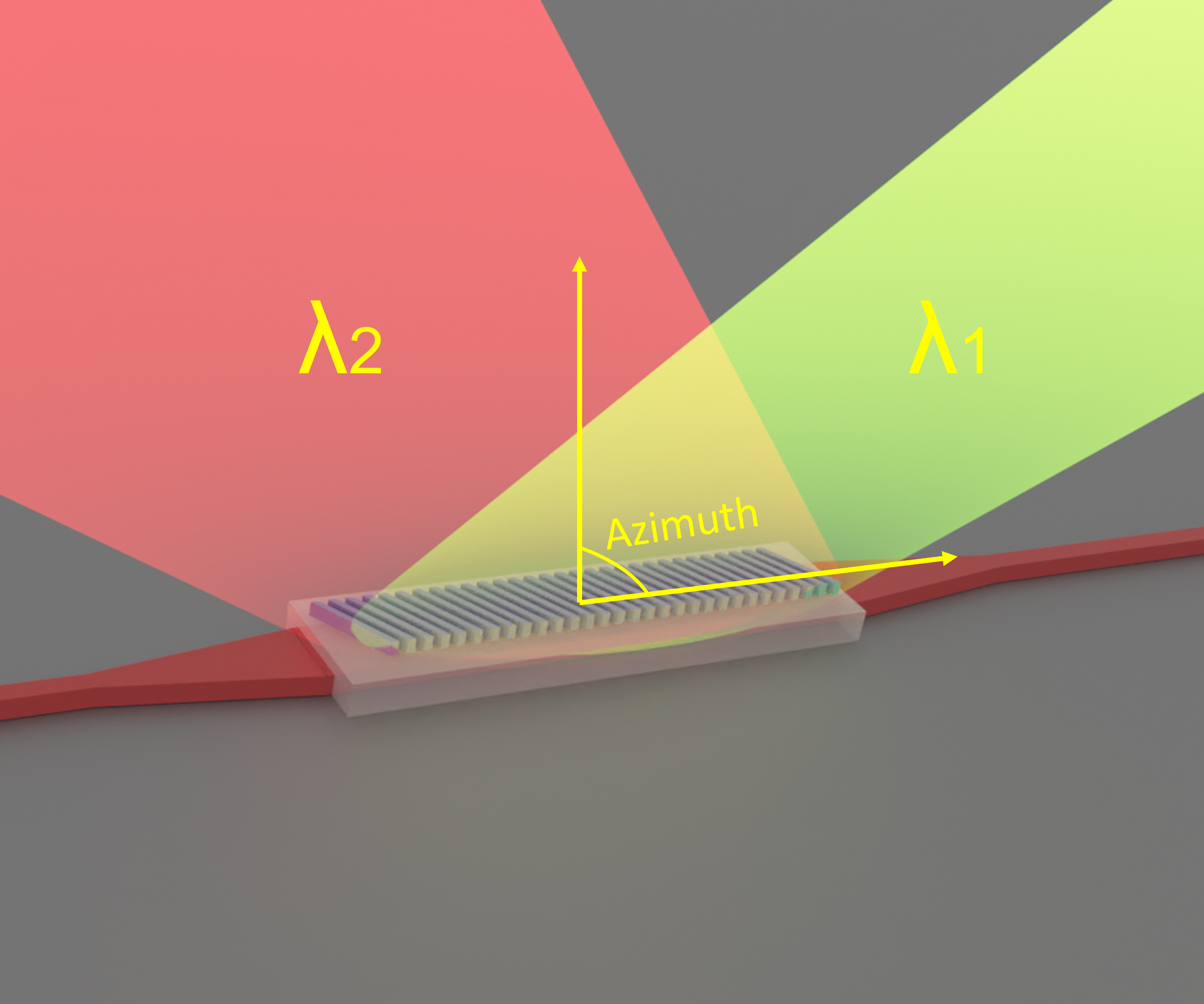}
  \caption{}
  \label{fig:3dwavelength:b}
\end{subfigure}
\caption{a) Illustration of radiation direction for different $\Theta$ slanting angles. b) Radiation direction for 
different $\lambda$ wavelength of light.}
\label{fig:3d_images}
\end{figure}

When altering the $\Theta$ slanting angleof the grating, the light propagates in the Zenith direction. Meanwhile, varying the wavelength $\lambda$ causes the light to change its angle in the Azimuth direction. These both properties are visualized in Figures \ref{fig:3dslant:a} and \ref{fig:3dwavelength:b} respectively.

The graphs below show the calculations of the angle changes from changing the wavelength and $\Theta$ slanting angle of the grating.

\begin{figure}
\centering
\begin{subfigure}{.5\textwidth}
  \centering
  \includegraphics[width=.8\linewidth]{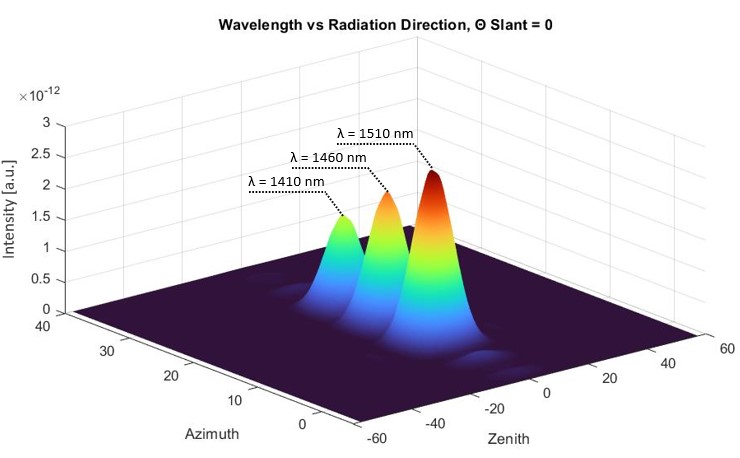}
  \caption{}
  \label{fig:annotated_wavelength:a}
\end{subfigure}%
\begin{subfigure}{.5\textwidth}
  \centering
  \includegraphics[width=.75\linewidth]{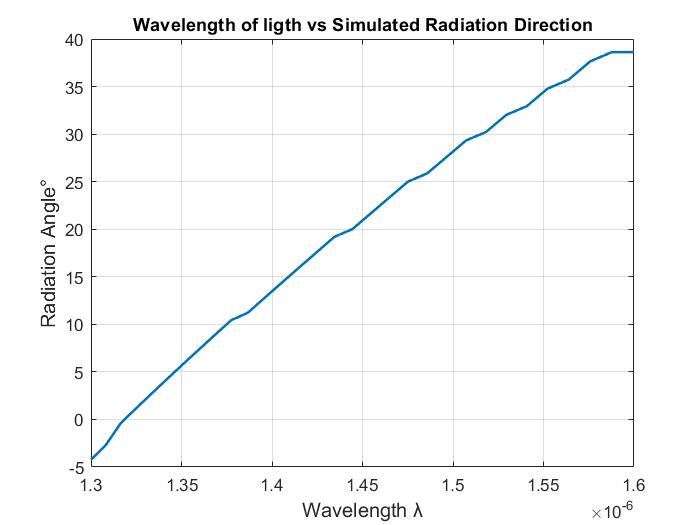}
  \caption{}
  \label{fig:annotated_wavelength:b}
\end{subfigure}
\caption{a) The far-field direction versus the $\lambda$ wavelength
of light in azimuth direction. b) Angle change versus the $\lambda$ wavelength of light.}
\label{fig:annotated_wavelength}
\end{figure}

Figure \ref{fig:annotated_wavelength:a} illustrates the change in far-field profile concerning the wavelength of light, revealing a noticeable shift in the azimuth direction. The far-field simulations were conducted with the grating slanting at 0$^\circ$, a filling factor of 0.5, 42 grating teeth, and a grating width of 10$\mu$m.

In Figure \ref{fig:annotated_wavelength:b}, it is evident that by adjusting the wavelength, light can be directed up to 42.8$^\circ$. Despite significant technological advancements, achieving a widely tunable broadband laser remains a challenging endeavor\cite{Coldren2022}.

\begin{figure}
\centering
\begin{subfigure}{.5\textwidth}
  \centering
  \includegraphics[width=.8\linewidth]{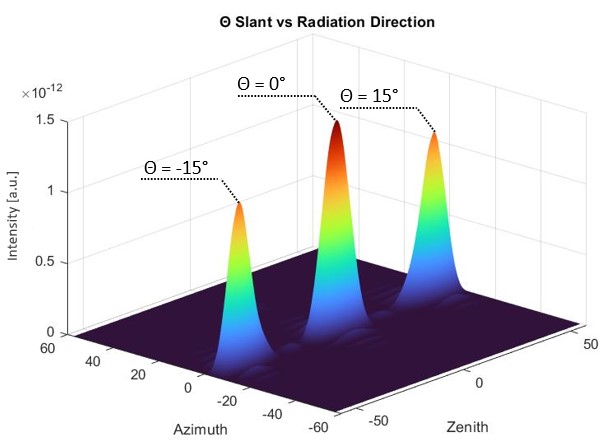}
  \caption{}
  \label{fig:annotate_slant:a}
\end{subfigure}%
\begin{subfigure}{.5\textwidth}
  \centering
  \includegraphics[width=.75\linewidth]{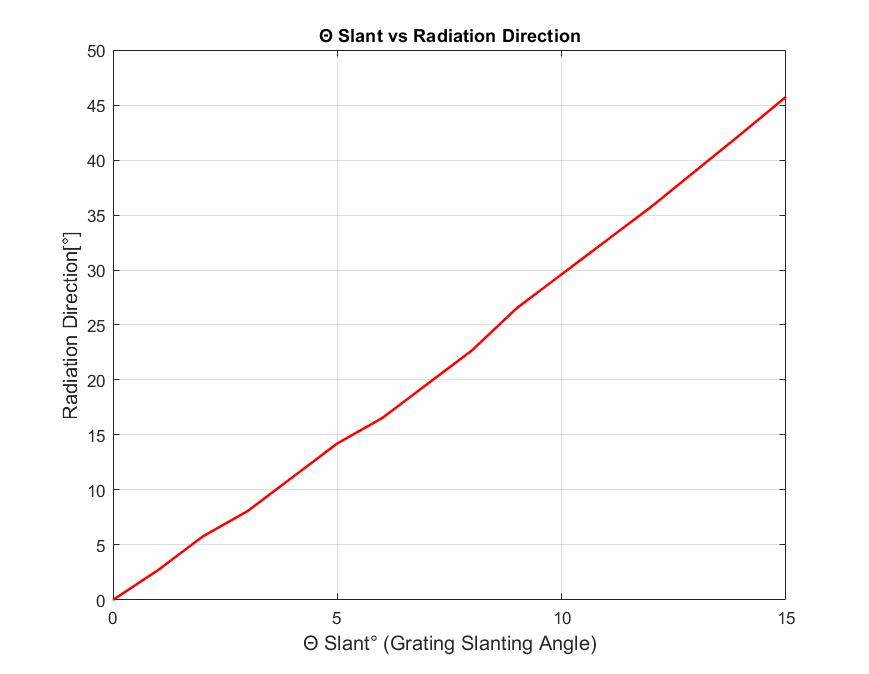}
  \caption{}
  \label{fig:annotate_slant:b}
\end{subfigure}
\caption{
a) The radiation direction for various $\Theta$ tilt angles in zenith direction.
b) Radiation direction for different $\Theta$ slanting angles of the grating coupler.
}
\label{fig:annotate_slant}
\end{figure}

In figure \ref{fig:annotate_slant:a}, the far-field pattern of the slanted GC is presented for two distinct slanting angles, one positive and one negative. Figure \ref{fig:annotate_slant:b} depicts the correlation between the GC tilt angle, denoted as $\Theta$, and the corresponding radiation direction. This observation not only highlights the capability to shift the radiation angle by tilting the gratings within the $\pm$15$^\circ$ range but also emphasizes the potential for achieving a broad field of view, with the span of  $\pm$45.75$^\circ$.

\subsubsection{Apodization}

The light within the gratings undergoes exponential decay. Preliminary simulations confirm this observation, as demonstrated by sweeping the length of the grating (refer to figure \ref{fig:side_view}) and subsequently extrapolating the values exponentially. 

\begin{figure}[h]%
\centering
\includegraphics[width=0.5\textwidth]{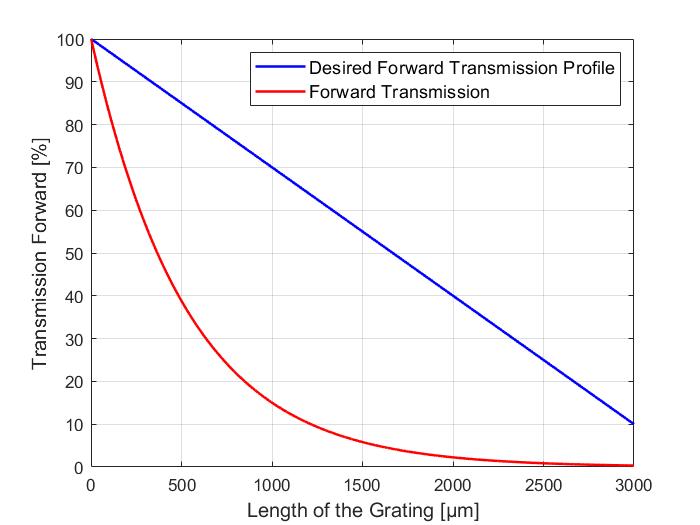}
\caption{The exponential decay of the light with the length of the grating coupler.}\label{fig:exponential_desired}
\end{figure}

However, for LiDAR applications, achieving high precision necessitates uniform illumination of the object. The simulated forward transmission profile of the grating versus its length is depicted below. 
The desired line in this context indicates the preferred pattern of change in forward transmission concerning the length.
Apodization involves adjusting geometric parameters such as periodicity and etch length along the grating coupler, aiming to achieve improved performance  \cite{Apodizationtheory}.

The apodization algorithm involves simulating a grating with 50 teeth, introducing light, determining the amplitude, and subsequently re-simulating the grating with the previously simulated amplitude as the input. The process includes calculating the amplitude and adjusting the filling factor accordingly in the course of the simulation. The wavelength of light is $\lambda = $1550 nm.

Following the simulations, the intended forward transmission profile has been successfully attained and is illustrated in the figure below. 
Figure \ref{fig:apodization} displays nine sections with varying filling factors, and the changes in these filling factors are visually represented on the same image.
\begin{figure}[h]%
\centering
\includegraphics[width=0.5\textwidth]{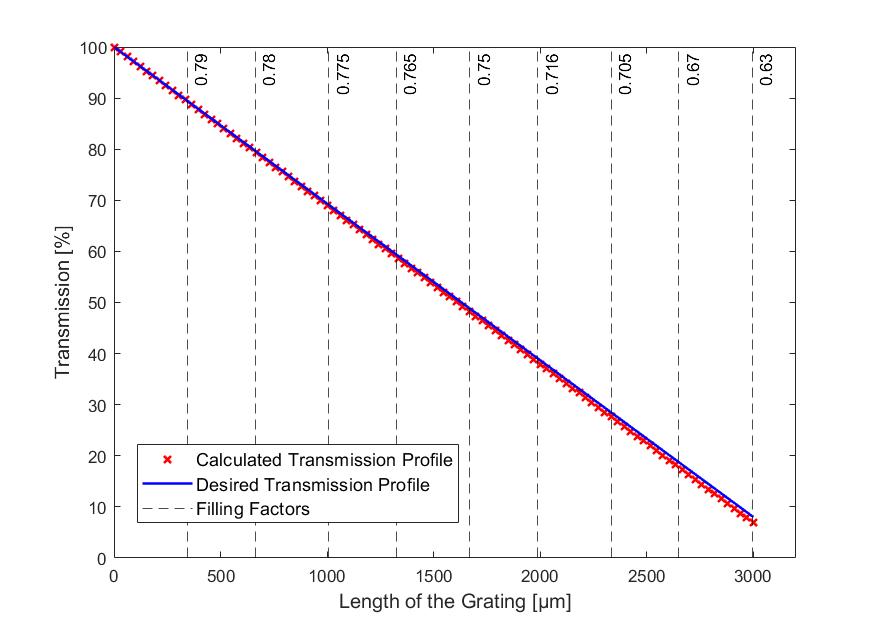}
\caption{Calculated intensity profile versus the desired line and filling factors at each section.}\label{fig:apodization}
\end{figure}

Following this, we can extract all the transmissions as depicted in the figure, presenting the simulated transmission. It is evident that the upward transmission is optimized to be the highest compared to the other transmission directions at the output.

\begin{figure}[h]%
\centering
\includegraphics[width=0.5\textwidth]{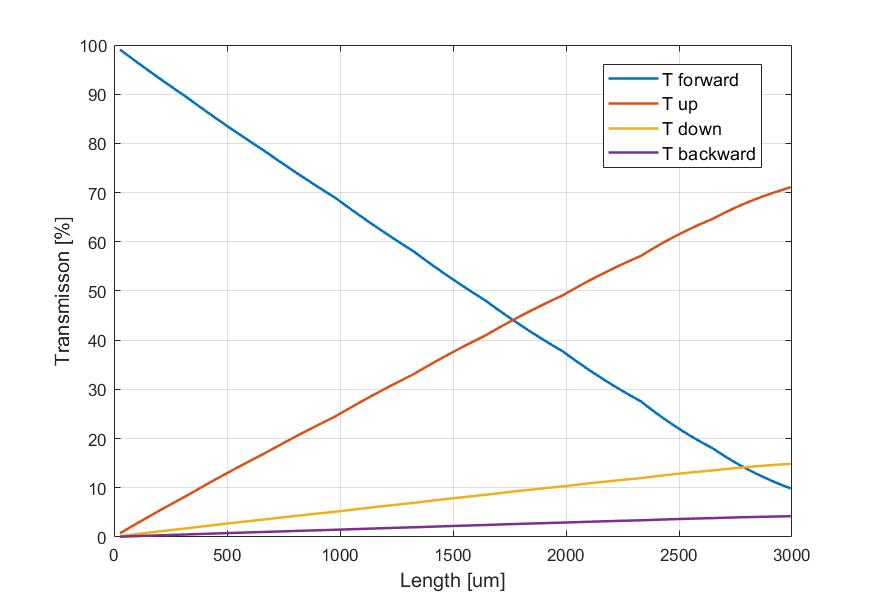}
\caption{All transmission profiles after the apodization.}\label{fig:allts}
\end{figure}

This maximizes the overall performance and sensitivity of the sensing system, leading to more reliable and precise measurements. Furthermore, employing a grating that spans 3mm length leads to a reduction in the Full Width at Half Maximum (FWHM) of the beam, consequently enhancing the resolution for sensing \cite{Wang2021}. The simulated Full Width at Half Maximum (FWHM) of the beam is illustrated in the figure below. It can be concluded that the FWHM of the beam in horizontal (azimuth) direction reaches 0.026$^\circ$ at a length of 3mm.

\begin{figure}[h]%
\centering
\includegraphics[width=0.5\textwidth]{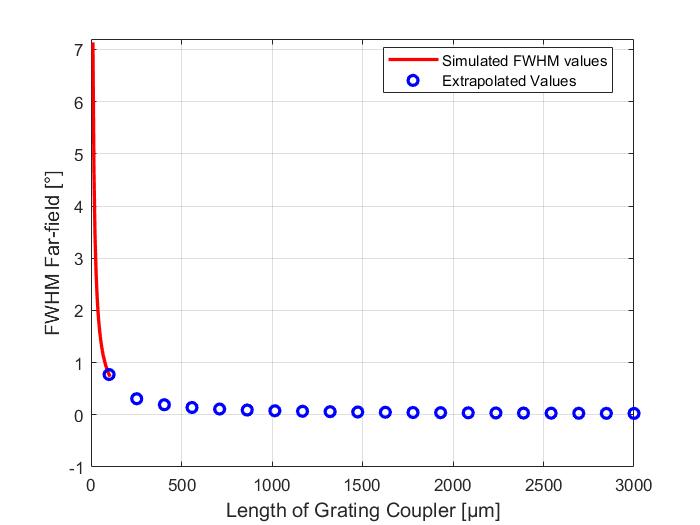}
\caption{FWHM change of the farfield vs the length of the grating}\label{fig:fwhm}
\end{figure}

\section{Experimental Results}
\begin{figure}[h!]%
\centering
\includegraphics[width=0.77\textwidth]{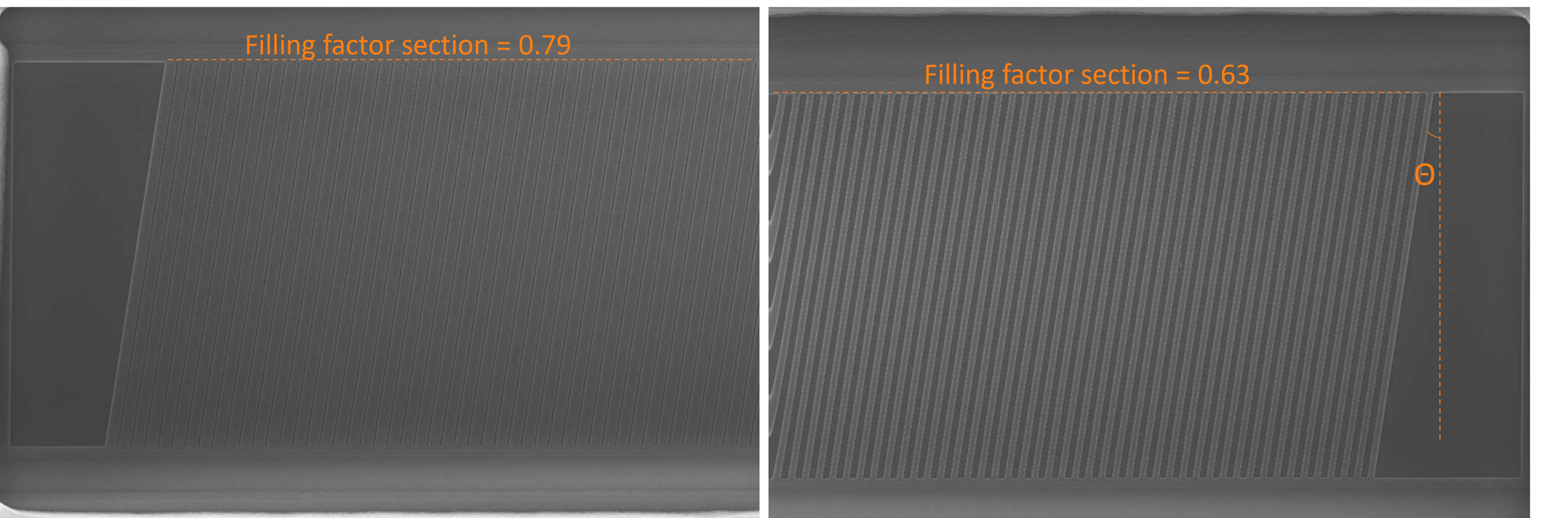}
\caption{SEM image of the fabricated grating coupler with the first and the last filling factor sections.}\label{fig:fabricatied_Grating}
\end{figure}

The grating is fabricated at SiPhotonIC ApS \cite{SiPhotonics} fast prototyping foundary in Denmark, Virum using Deep UV lithography on standard SOI platform but with the additional layers of $\text{Si}_3\text{N}_4$ and $\text{SiO}_2$.
 The figure \ref{fig:fabricatied_Grating} shows the fabricated grating coupler with different apodizations and the $\Theta$ slanting angle can also be seen. In the figure is shown that the grating teeth change from being thin to thick. The width of the grating is 20$\mu$m.

\begin{figure}
\centering
\begin{subfigure}{.5\textwidth}
  \centering
  \includegraphics[width=1.1\linewidth]{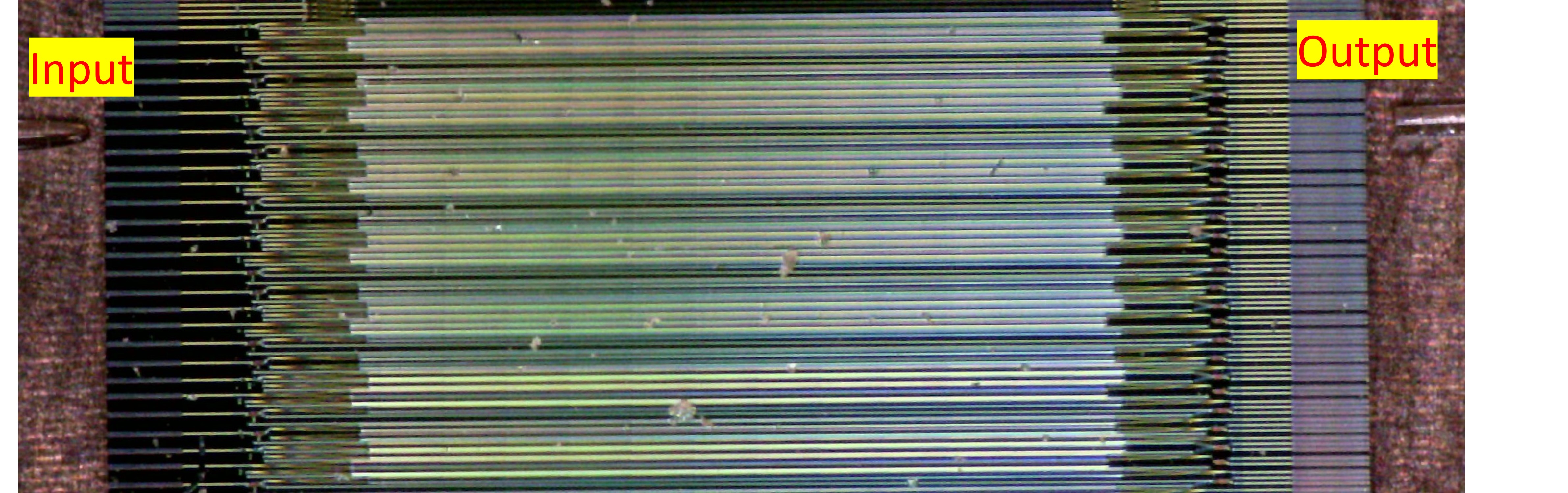}
  \caption{}
  \label{fig:test:chip}
\end{subfigure}%
\begin{subfigure}{.5\textwidth}
  \centering
  \includegraphics[width=.66\linewidth]{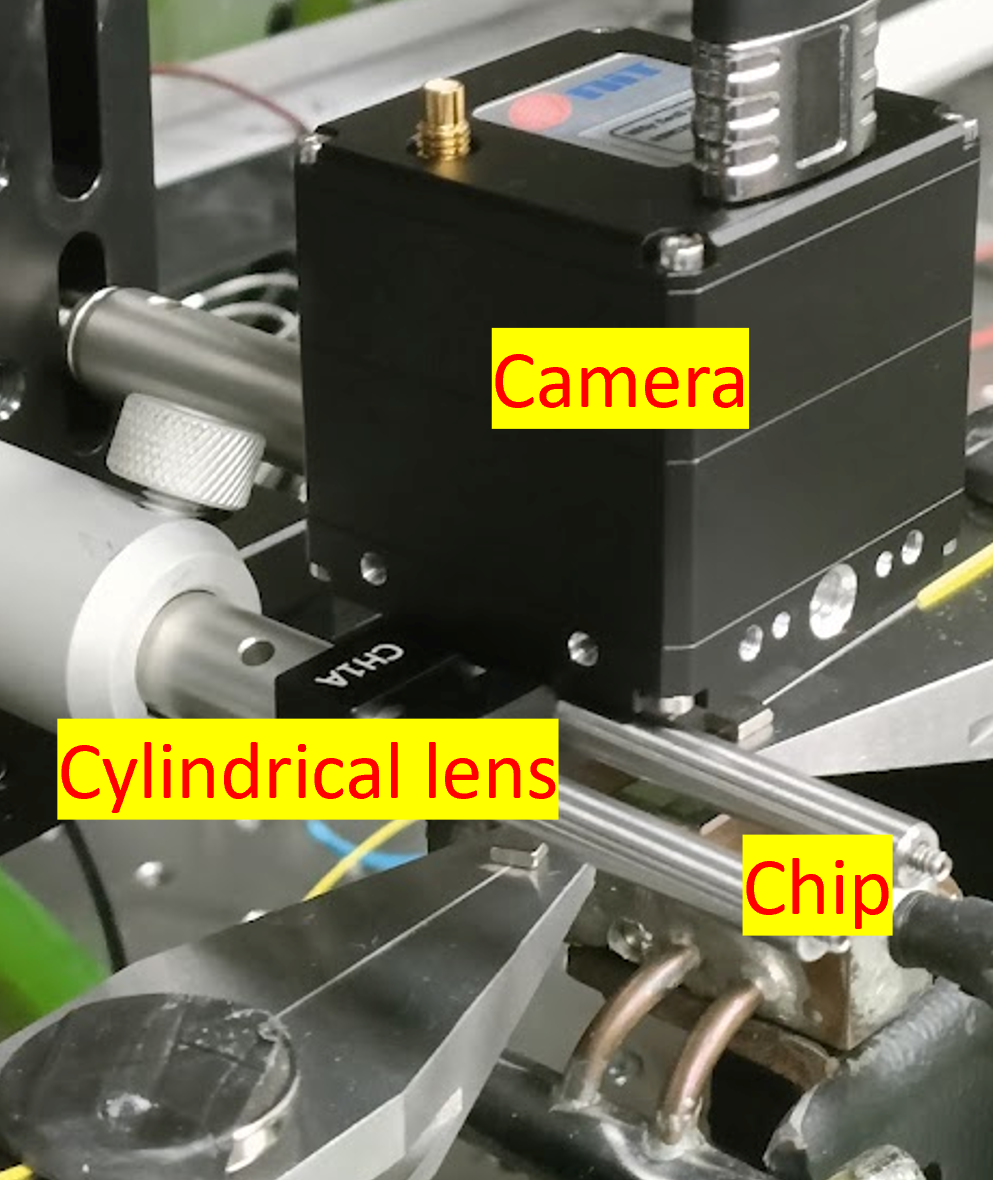}
  \caption{}
  \label{fig:test:camera}
\end{subfigure}
\caption{a) Fabricated integrated photonic circuit with slanted grating arrays with different $\Theta$ slanting angles. b) Characterization setup with SWIR infrared camera and the cylindrical lens.}
\label{fig:test}
\end{figure}

The fabricated photonic integrated circuit is shown in the figure \ref{fig:test:chip}. The grating with different slatnings can be seen, the designed gratings are 3mm long. 
The SWIR (Short-Wave Infrared) camera with resoltuion of 320x256 was implemented with the primary objective of characterizing light beams. For the experimental results, a tunable laser with a wavelength span of 30 nanometers, ranging from 1540nm to 1570nm was used.
The lensed fiber is used to couple the light into the edge couplers  while for the collection a cleaved fascet single mode fiber is used. 
The resulting radiation pattern is observed on the SWIR camera's sensor. To collimate the light on the chip, a cylindrical lens is employed, as illustrated in figure \ref{fig:test:camera} which is placed at a distance of 24.4 mm from the photonic circuit.

\begin{figure}
\centering
\begin{subfigure}{.5\textwidth}
  \centering
  \includegraphics[width=.75\linewidth]{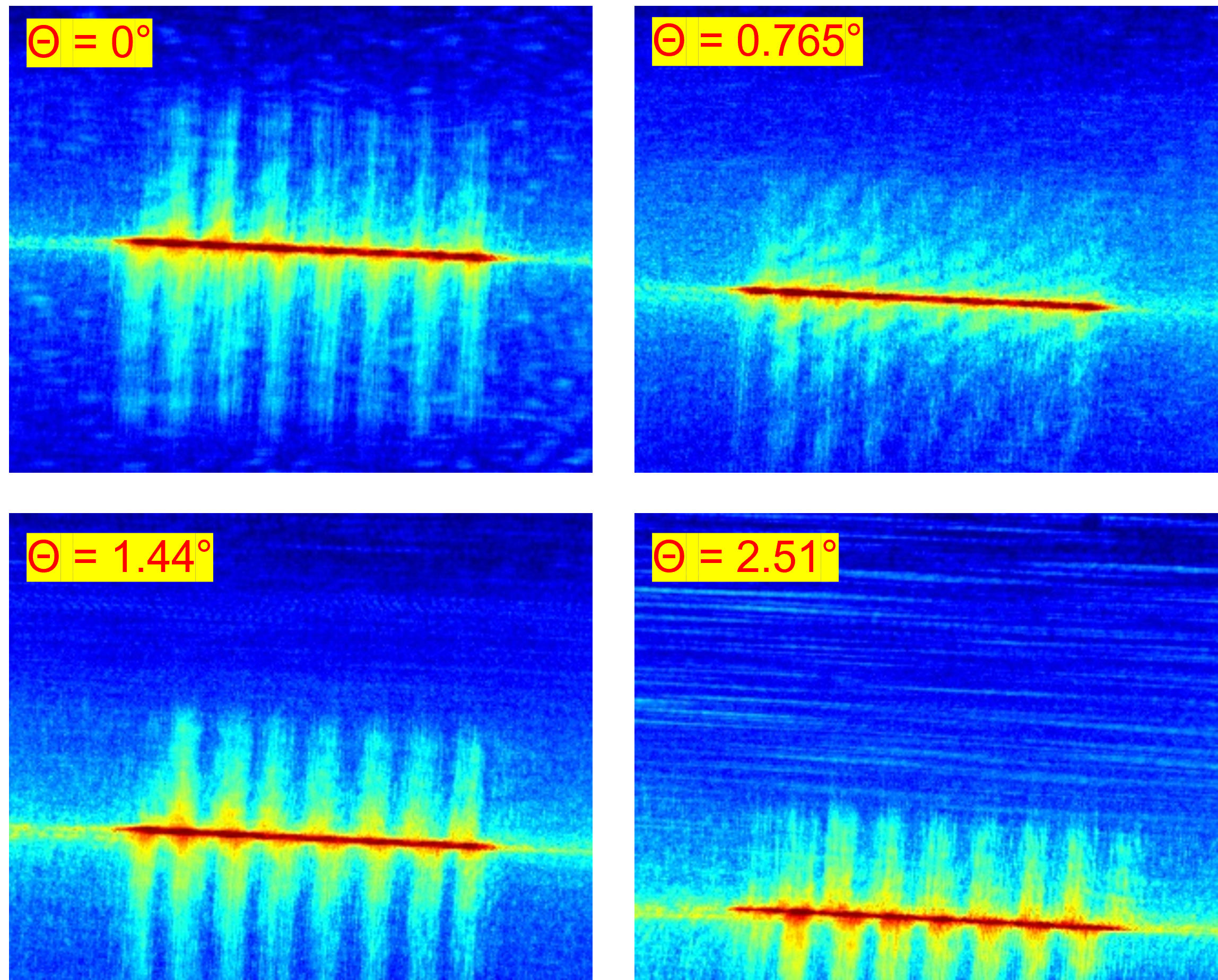}
  \caption{}
  \label{fig:slant_exp:cal:infrared}
\end{subfigure}%
\begin{subfigure}{.5\textwidth}
  \centering
  \includegraphics[width=.8\linewidth]{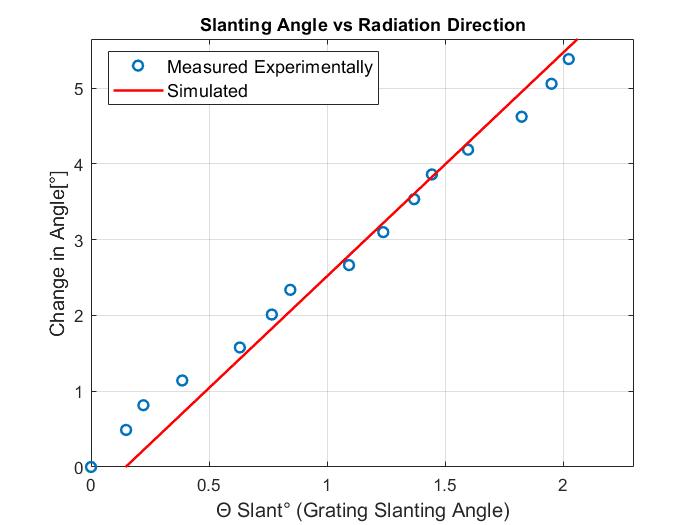}
  \caption{}
  \label{fig:slant_exp:calc}
\end{subfigure}
\caption{a) Infrared camera image depicting the angle change resulting from different $\Theta$ slanting angles. b) Comparative analysis between calculated and simulated angle changes with respect to $\Theta$ slanting angles.}
\label{fig:slant_exp}
\end{figure}

The setup enabled the detection of radiation angles and, most critically, the profiling of beams from the gratings. It was established that the actual beams and their uniform profiles closely matched the outcomes from the simulation, affirming the reliability of the initial calculations. Additional experiments were conducted to investigate the effects of varying grating slant angles and different wavelengths on the radiation angles, as depicted in the figures \ref{fig:slant_exp:cal:infrared} and \ref{fig:slant_exp:calc} respectively.

\begin{figure}
\centering
\begin{subfigure}{.5\textwidth}
  \centering
  \includegraphics[width=.75\linewidth]{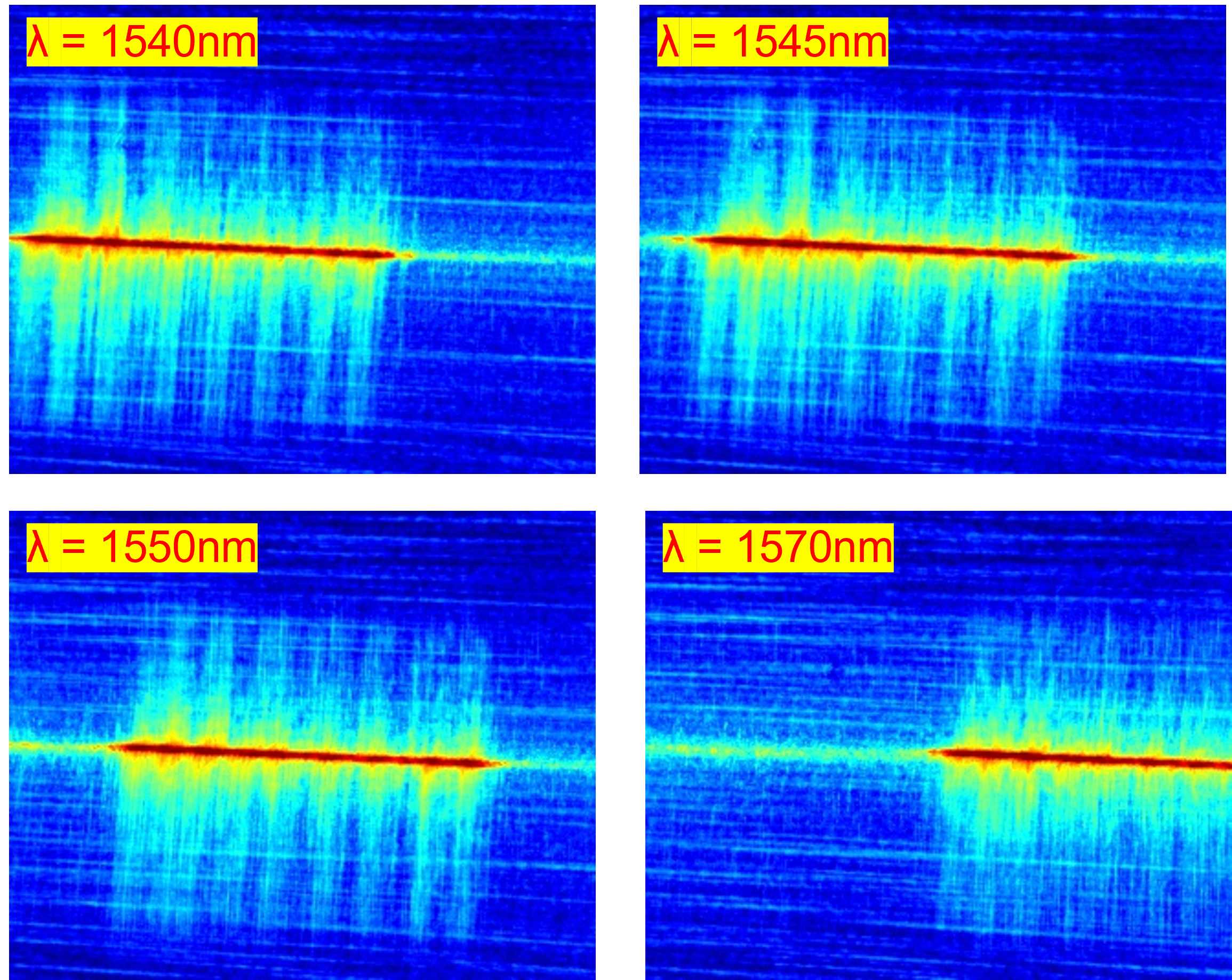}
  \caption{}
  \label{fig::wl_exp:infrared}
\end{subfigure}%
\begin{subfigure}{.5\textwidth}
  \centering
  \includegraphics[width=.8\linewidth]{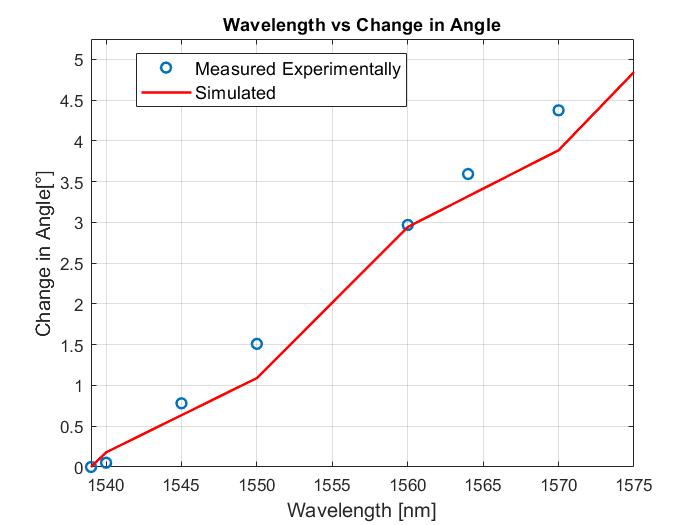}
  \caption{}
  \label{fig:wl_exp:calc}
\end{subfigure}
\caption{a) Infrared camera image depicting angle variations due to different $\lambda$ wavelengths of light. b) Comparative analysis between calculated and simulated angle changes with respect to the $\lambda$ wavelength of light.
}
\label{fig:wl_exp}
\end{figure}

Finally the comparison of the measured and the calculated angles have been concluded which can be seen in the following figure. 

Moreover, we can see that the emission profile from the grating is exactly 3mm which shows that the apodization for the grating has been done correctly. 

\begin{figure}[h!]%
\centering
\includegraphics[width=0.37\textwidth]{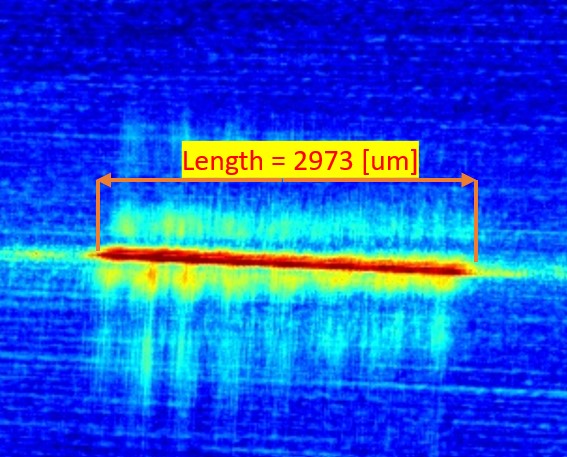}
\caption{Calculated length of the radiated light from the grating as seen with the infrared camera}\label{fig:length}
\end{figure}

However, it's important to note some limitations in our measurements. The use of a tunable laser with a span of 30nm restricts the full exploration of wavelength-dependent effects, and the absence of concise mechanical alignments hinders a comprehensive mapping of emission angles with larger slanting gratings. The presented plots serve as an initial assessment, demonstrating the alignment of calculated and experimental slopes. Future measurements with a broader wavelength span and improved mechanical alignments are necessary for a more comprehensive evaluation.

\section{Conclusion}

In conclusion, our study introduces a passive steering concept for LiDAR systems using slanted grating couplers. The theoretical calculations predict a steering capability of 91.5$^\circ$x42.8$^\circ$.
Strategic use of the layerstack enhances the transmission upward while reducing the transmission that goes into the substrate. By channeling the emitted light primarily in the upward direction, we ensure that the LiDAR system receives a cleaner signal. This enhancement in signal is crucial for accurate and reliable detection and ranging capabilities, which are fundamental aspects of LiDAR technology.
Furthermore, the integration of 3mm-long gratings  contributes to FWHM of the farfield at 0.026$^\circ$ in horizontal (azimuth) direction, enhancing overall sensing resolution while also maintaining a linear emission pattern due to the apodization of the grating.  
This work lays the foundation for the development of efficient and economical LiDAR systems, paving the way for broader applications in diverse optical systems.
While our experimental results demonstrate promising alignment with theoretical calculations, it's crucial to acknowledge limitations. The use of a tunable laser with a limited span and the absence of precise mechanical alignments impose constraints on our measurements. To achieve a more comprehensive evaluation, future experiments will incorporate a broader wavelength span and improved mechanical alignments. Additionally, exploring receiving characteristics of the grating couplers from various angles is essential for advancing towards the development of a fully functional LiDAR system.

\section*{Declarations}

\begin{itemize}
  \item \textbf{Funding:} We wish to acknowledge the support received from H2020-MSCA-ITN-2019-DRIVE-IN under Grant No. N860763.

  \item \textbf{Conflict of interest/Competing interests:} The authors declare that they have no competing interests.

  \item \textbf{Ethics approval:} Not applicable for this study as it does not involve human subjects.

  \item \textbf{Consent to participate:} Not applicable for this study as it does not involve human subjects.

  \item \textbf{Consent for publication:} Not applicable for this study as it does not involve human subjects.

  \item \textbf{Availability of data and materials:} For inquiries regarding the availability of materials, please contact the corresponding author.

  \item \textbf{Code availability:} Not applicable for this study.

  \item \textbf{Authors' contributions:}
    \begin{itemize}
      \item Vahram Voskerchyan: Research and development of the system.
      \item Yu Tian: Layout design.
      \item Francisco M. Soares: Supervision.
      \item David Álvarez Outerelo: Provision of laboratory equipment and resources.
      \item Francisco J. Diaz-Otero: Project management.
    \end{itemize}
\end{itemize}


\end{document}